# Dream Formulations and Deep Neural Networks: Humanistic Themes in the Iconology of the Machine-Learned Image

Emily L. Spratt [1,2*]

[1]Department of Art and Archaeology, Princeton University, Princeton, NJ
[2]Research Department, The Frick Collection, New York, NY
[*]Correspondence should be addressed to emilylspratt@gmail.com

**Abstract: This paper addresses the interpretability of deep learning-enabled image recognition processes in computer vision science in relation to theories in art history and cognitive psychology on the vision-related perceptual capabilities of humans. Examination of what is determinable about the machine-learned image in comparison to humanistic theories of visual perception, particularly in regard to art historian Erwin Panofsky's methodology for image analysis and psychologist Eleanor Rosch's theory of graded categorization according to prototypes, finds that there are surprising similarities between the two that suggest that researchers in the arts and the sciences would have much to benefit from closer collaborations. Utilizing the examples of Google's DeepDream and the Machine Learning and Perception Lab at Georgia Tech's Grad-CAM: Gradient-weighted Class Activation Mapping programs, this study suggests that a revival of art historical research in iconography and formalism in the age of AI is essential for shaping the future navigation and interpretation of all machine-learned images, given the rapid developments in image recognition technologies.**

**N.B. This paper is published in the art historical journal Kunsttexte.de.**

In 2015, Google released a computer vision program that aids our understanding of the mechanisms at work in machine-based image classification.[1] Entitled DeepDream for its association with a type of artificial intelligence called deep learning, and also a reference to the unconscious processes that allow the brain to create dreams, the program produces a visualization of the image-recognition process (fig. 1). Given the remarkably vivid visual effects that the program generates, DeepDream has quickly been adopted for use as an artistic tool to augment existing images, a phenomenon that has furthered the algorithm's popularization, yet also has contributed to the media's misinterpretation of its original function.[2] In this paper, the use of this program, among others, as a tool to better understand image recognition—as it was initially intended—will be examined in relation to theories on the vision-related perceptual capabilities of humans and machines.

Computer scientists often employ anthropomorphizing metaphors such as "deep dream" in their research, misleadingly projecting the possibilities and limitations of the human condition onto the mechanisms that empower the computational abilities of machines.[3] Although nothing on the programming level of the DeepDream algorithm actually replicates neurobiological processes, this term and others like it are indicative of the development of artificial intelligence to emulate aspects of human perception.[4] Vision technology, in particular, can be directed in ways that interpret the world analogously to humans.[5] It is a burgeoning area of research that is not only transforming computer science and robotics, but is offering interesting possibilities for the augmentation of human visual perception, our understanding of the underlying mechanisms at work in machine-guided image recognition, the creation of new tools for artists, and the ability to produce imagery.[6] Given these astonishing developments in computer vision science that challenge our traditional understanding of both human- and machine-based visual perception and their creative potential to produce visual content, this area is ripe for philosophical investigation.

Google's DeepDream project exemplifies the remarkable growth of computer vision science in the last few years, and advances in image-recognition technology at large have already led to results that are comparable to the performative aspects of the interpretation of visual information through human sight itself.[7] Indeed, the more computer vision technology is able to mimic human visual processes, the better its real-world applications can be guided, and this is particularly the case with image recognition and retrieval tools. While research in this area is often sequestered to the fields of computer science and business development in the technology sector, I aim to demonstrate how the analysis of images with deep-learning techniques can engage the humanities, complement existing sociocultural theories, and offer the possibility of new methodologies for image analysis that take cognitive psychology into consideration.[8] Let us therefore begin our analysis of the development of DeepDream as a part of the larger turn toward deep-learning techniques in computer vision science in order to examine some ways in which images are processed by machines in comparison to humans. In essence, let us consider the iconology of the digital image vis-à-vis deep neural networks, what I have termed the machine-learned image.[9]

    As a part of the 2014 ImageNet Large-Scale Visual Recognition Challenge, a deep convolutional neural network architecture was created by a team of computer scientists from Google, the University of North Carolina at Chapel Hill, the University of Michigan at Ann Arbor, and Magic Leap Incorporated.[10] The goal of this competition was to improve the classification of images and the detection of their contents. The success of this research group's network was based on its computational efficiency and enhanced ability to analyze images through multiple wide layers of coding. Applied to the interpretation of two-dimensional digital images, this branch of machine learning, based on neural networks, more commonly known as deep learning, can be imagined as a complex data filtration system that processes information layer by linear- and non-linear-algorithmic layer. Like a water filtration system that distills liquid by outputting only what has already been filtered to each subsequent level, a convolutional neural network has the capacity to filter information according to multiple parameters, the size of

its constituent parts being just one component of the system. In both examples, there are many steps in between the input and output, which allows for increased complexity in the final analysis.

While the traditional programming model is based on breaking down large problems into smaller solvable tasks, deep learning allows the computer to find its own solutions to presented problems through multiple interlaced layers. Although the programmer sets the initial parameters and provides the visual training set that a computer will utilize to detect certain images and their constituent parts, how this is actually achieved remains rather mysterious to the computer scientist.[11] Consider the development of facial recognition programs, such as the one created by Facebook: a system can learn to detect a face and compare it to other faces through layers of convolutional neural networks if a large and coded dataset from which the computer is trained is available. Untangling what the computer finds to be most predictive in facial recognition out of the many strengthened neural pathways that are found in the image-recognition process, however, is not always possible.[12] This unknown dimension, when the computer essentially becomes autodidactic, is called the "black box" of computer learning. Determining what the computer has created to optimize efficiency in its recognition of labeled and unlabeled images is a current trend in computer vision science, and reverse engineering what the computer has come up with is no easy task.

DeepDream is one attempt to uncover what occurs in the black box through visualization. Led by Google engineer Alexander Mordvintsev, the DeepDream team harnessed aspects of the deep neural network architecture developed by the 2014 Image Challenge research group and visualized what was going on in the network as it processed information.[13] Recognizing that neural networks trained to discriminate between different types of images also have the information needed to generate them, the Google team aimed to use visualization as one way to determine whether the network had correctly learned the right features of an image by displaying their associations in the classification process. For example, a neural network designed to recognize dumbbells seemed to be doing so correctly, but was in fact inaccurately making the classification. Since the computer was trained with numerous images of weightlifters holding dumbbells (fig. 2), the neural net determined that the weightlifter's arm was an essential part of the classification of the object since the majority of online images of dumbbells include an arm curling them.[14] For this reason, the DeepDream algorithm generated a visualization of dumbbells with an attached arm (fig. 3)! DeepDream thus gives us visual information that can not only correct these types of training errors, but can also tell us more about the classification process.

Another research project, developed as an open-source cloud platform at the Machine Learning and Perception Lab at Georgia Tech and entitled "Grad-CAM: Gradient-weighted Class Activation Mapping," allows us to see what convolutional neural networks determine to be the most predictive locations in an image to identify its content, and creates text labels of its findings that classify the image's content, or captions that describe the image in further detail.[15] Like DeepDream and other programs, Grad-CAM allows us to peak into the black box of machine learning with visualization techniques that reveal components of the image-recognition process, yet adds semantic descriptions to its finding. Both projects are significant across academic fields and have particular value for the museum and cultural heritage management world, as this research provides much insight on ways to improve machine-based image classification. The Frick Collection is one museum where this type of technology is sought, and plans there are under way to begin collaborations with computer scientists for its development for application to art historical materials.[16]

      While Grad-CAM's program demo highlights its use with images of contemporary life that are mostly open-source and randomly generated, media-focused photography, I applied the program for use in art history and ran a selection of images from the collection of the Metropolitan Museum of Art through it.[17] In complex images such as El Greco's *Christ Healing the Blind,* the program labeled the image as a carousel and identified places on the picture field outside of the central scene of Christ's performance of the miracle as the locations that were significant for identification (fig. 4). By contrast, an image of a thirteenth-century icon depicting the Madonna and Child is labeled by Grad-CAM as a "book jacket" and highlights the Virgin's face and the surrounding area, but does not emphasize Christ.[18] Late Antique parchment leaves from Egypt displaying Greek text were identified as "honeycomb," and the program selected seemingly random locations including the background mat on which the artifact rests as significant for its content.[19] Not surprisingly, a photograph of Georgia O'Keeffe clutching her coat collar, by Alfred Stieglitz, performed somewhat better with the program (fig. 5). It recognized the hands on the garment as the significant part of the image when generating a caption, but the text describes the image as "a man is holding a cell phone in his hand."[20]

      Like DeepDream, Grad-CAM's results are highly biased according to its data training set, and the results of its analysis for types of images that it is not familiar with reflect this. The input of images reflecting different types of media erroneously produce the suggestion that a different object type is being analyzed, rather than picking up on the displayed formal content, as if all images are photographs of things found in the contemporary word. The flatness of the icon

suggests the advertisement-like nature of a book cover, the El Greco painting's multiple eclipsing scenes is classified as a carousel, parchment betraying its deterioration over time is seen as honeycomb, and a photo from 1918 is given a caption that assumes the invention of communications technology used today. In future research it would be interesting to train the program using art historical images with additional parameters and to compare the results.

  Both the DeepDream and the Grad-CAM program demonstrate how machine-based image classification can be compared to the human perception of an object through one's association of it with known examples of its type. Generally stated, the machine, not unlike a human, learns to recognize a given image because it has seen it or something close to it before. If based on a training set that has multiple contemporary images of people with their mobile phones, the program assumes that grasping hands would hold a phone, just as a child who did not know a world before such devices might assume they are in photographs of historical persons. For both the machine and the human, we don't know the precise mechanisms at work when an image is being translated into its higher-level meanings and associations, but we are certain that what has been seen and interpreted before plays a role in our visual understanding of newly presented things.[21]

  A major problem in image recognition for both machines and humans is the ability to accurately generalize and specify categories of objects.[22] Let's first consider this problem for machines. Since the internet is full of images of animals, they are overly represented in machine-learning training sets and are the usual examples that computer scientists employ to discuss classification issues. In the case of dogs, vision technology easily interprets a picture of a Golden Retriever as a dog but may incorrectly classify a Bedlington Terrier as a lamb or a Bichon Frise as a polar bear cub, since those breeds do not at first glance appear to be canine (fig. 6). It is important to note, however, that the probability of vision technology programs correctly interpreting dog breeds is higher than it would be for a category of objects such as medieval *encolpia*, since most available datasets are overrepresented in categories like dogs rather than in religious artifacts. In either case, how a machine can make these designations is even more difficult than it is for humans.

At a basic level the computer differentiates one image from the next according to the images it has been trained to recognize. One way it does this is by isolating the optimal and measurable geometric features of each object in the picture and then comparing them to the object measurements found in the labeled dataset. From this training, the machine develops a distribution of probabilities for each pixel of an image to indicate the most likely locations in the picture plane where the object will appear. As already exemplified with Grad-CAM, with knowledge of the likely locations of pixels in the representation of a given object, it is possible to make maps based on the training data to see the relative probabilities of pixels falling into the locations we associate with a certain object and then assigning a percentage-based evaluation of the likelihood a given object is a certain type of dog. In essence, this is the naïve Bayes method of classification and is but one type of image-recognition solution. How we can interpret objects presented in an image at any angle with any background is quite complex, but is an essential classification problem that affects all forms of image-recognition techniques.

If we can algorithmically distinguish dog breeds using methods beyond Bayes's model in neural networks, we have the ability to train a neural network to recognize any class of objects. The more these pathways are enforced by using the network, the more accurate these predictions become. Machine-learning techniques are therefore incredibly useful for image recognition, but require large amounts of data to train upon in order to function with precision, and these analyses are limited in their formalistic interpretation of the data, whether it is labeled or not. It cannot be stressed enough how critical the dataset that the program is trained on is for the process of image recognition, and a large and vetted dataset that is specified for the image-recognition task at hand is usually not easily available.

These computer-recognition issues have some similarities to theories regarding the underlying mechanisms of human visual perception. While there is a long history of theories on human visual perception, development of the ability to categorize objects is one area of research that was of particular interest to psychologists in the last century.[23] In the 1970s, cognitive scientist Eleanor Rosch and others devised a theory of graded categorization according to prototypes in place of the Aristotelian, definition-based model of object categorization and in revision of Jean Piaget's pioneering work on cognitive development in children.[24] According to this theory, some components of a category are more significant than others. Thus, recognizing the similarity of an object to one's notion of its prototype is what cognitively leads to its identification. In this respect, a Golden Retriever really is a better example of a dog than a Bedlington Terrier or a Bichon Frise, or, to use the classic examples from Rosch's study, a robin is a more exemplary bird than a penguin, and a chair is more closely associated with the category furniture than is an ottoman.[25]

Our willingness to accept these assessments relies on the inherent biases we have in interpreting the relationships that types of things have with their prototypes.[26] According to prototype theory, these relationships are even measurable according to the response time required to recognize the image at hand.[27] Not surprisingly, non-prototypical things require more time to register than objects that seem to organically represent their type. The theory is also tied to developmental psychology, as children explore the boundaries of prototypes by over-identifying them, such as their description of all four-legged creatures as dogs.[28] Nonetheless, the extensive labeling of things in relation to their perceived prototype is likely an important part of the cognitive formation of different types within the category as well.

More than sixty years after the first developments in machine learning, computer scientists still consider image-based AI to be loosely analogous to a child's developmental state. In the case of general versus specific image classification issues, this really does seem to be the case, even if so only superficially. Since DeepDream was trained using an online image set that consists of an overly represented number of animals, the network biased the classification of all the objects it analyzed as animals, especially as dogs, and thus visually manifested them (fig. 7). The Google team also observed other tendencies in the classification process, such as the appearance of towers and pagodas on horizon lines.[29] This phenomenon may be interpreted like the example of the dumbbell with the attached arm. Since the images that trained the network included landscapes, which more often than not featured a building on the horizon line, the computer created its classification of a horizon line to include an architectural structure punctuating it (fig. 1). This information tells us that much work remains to be done in general versus specific machine-based image classification vis-à-vis the DeepDream program. It also demonstrates the power of visualizing the image-recognition process to understand it better, and the inherent visual relationships between the images from the training dataset with the analyzed image.

Computer vision science reminds us that our own visual abilities in identifying images are also the result of training and are rarely unbiased. How we perceive an object, classify it, and interpret it (in relation to other objects and other images) is a complex question that may not be trendy to ask in art history today, but has everything to do with the origins of the discipline, and has particular relevance to the way art historians in pre-Modern fields often approach the objects they study. It was this very question that drove Erwin Panofsky to make a science of identification for art history that is still used today.[30]

In order to achieve a higher-level understanding of Renaissance art, Panofsky posited three formal and empirically based stages of analysis that themselves progress from part to whole and then back again. This methodology was greatly elaborated in *Studies in Iconology: Humanistic Themes in the Art of the Renaissance*, which was first published almost eight decades ago.[31] What the DeepDream algorithm and programs like Grad-CAM reveal about the nature of image classification using neural networks applies to Panofsky's iconographic schema and fits within the stage of analysis he termed "primary" or "pre-iconographic." Nodding to Heinrich Wölfflin, Panofsky called this stage "pseudo-formal analysis" and elaborated on the mechanisms at work in a visual description. He theorized that familiarity with previously experienced objects and events was the mental equipment necessary for interpretation at this level and that insight into the history of style could be used to control this sort of analysis.

Seemingly ahead of his time as far as cognitive psychology is concerned, Panofsky was actually importing the bridge already built by Ernst Cassirer between the study of human perception and issues in classification to the field of art history.[32] In his research, Panofsky was particularly sensitive to the role that pre-existing biases play in the interpretation of art and how they could be balanced with what he differentiated as the history of style, the history of types, and the history of cultural symbols. In "A Neural Algorithm of Artistic Style," Leon A. Gatys and others reach an astonishingly similar conclusion in their research on deep neural networks and the ability of algorithms to replicate established artistic styles. They state: "All in all it is truly fascinating that a neural system, which is trained to perform one of the core computational tasks of biological vision, automatically learns image representations that allow the separation of image content from style. The explanation could be that when learning object recognition, the network has to become invariant to all image variations that preserve object identity. . . Thus, our ability to abstract content from style and therefore our ability to create and enjoy art might be primarily a preeminent signature of the powerful inference capabilities of our visual system."[33]

Unfortunately, advances in computer vision science which demonstrate the exploration of deep-learning techniques for image-recognition tasks are typically first encountered by art historians in the mainstream media, which too often inadequately convey the use/value of the development of these new digital tools. In the case of DeepDream, the popular press immediately seized upon the mesmerizing visualizations produced by the algorithm and incorrectly explained them as computer hallucinations or unconscious visual manifestations that could be seized upon by artists. This in turn led to Google's release of the algorithm so anyone could process their images and receive the visual results, which were later even made customizable.[34] That the visualization would be turned into an augmented paintbrush was an unintended, yet interesting, consequence of the research that shifted the focus of DeepDream away from its purpose of exposing the processes of the neural network during image classification.

By spring of 2016, an art cooperative in San Francisco called the Gray Area Foundation was able to hold a moderately successful benefit featuring a group of artists who were using DeepDream for image enhancements. At this point, the algorithm was so misinterpreted that a science editor from the Washington Post, commenting on the DeepDream excitement, was able to comfortably make the suggestion that the auction sales represented the art market's endorsement of the use of computer science and AI in contemporary art rather than an example of Silicon Valley philanthropy.[35] Unfortunately, the kitsch-like quality and mainstream hype about the visualizations have not helped the algorithm to be considered more seriously for its original purpose, nor has it improved the stigma of "computer art" in the art market.[36]

Fortunately, the initial research intentions of the DeepDream project have continued to be addressed at Google through the research of Chris Olah, Alexander Mordvintsev, and Ludwig Schubert in response to the scientific community's recognition that neural networks require interpretability to best harness and direct their use/values.[37] While interest in understanding what occurs in the black box has been prevalent, the ability to now directly address this issue in computer vision science truly marks another milestone for artificial intelligence, and has recently led to the development of what some researchers are calling the emergent field of neural network interpretability. Indeed, the developments in computer vision science are occurring so quickly that it is difficult for collaborators from the humanities to plunge into research projects with computer scientists and to stay apace given the different timelines required by each domain in all aspects of the research process.[38] This is not surprising given the faster rates in which technological developments occur in science in comparison to the time required for analytical responsiveness to sociocultural phenomena in the arts, and how the corresponding infrastructures that are geared to support knowledge production in each domain end up emphasizing these differences.

Unlike the Grad-CAM program, which addresses issues in what computer scientists term the *attribution* of an image and define as the correspondence between parts of the picture plane to its network activation, the Google project focuses on *feature visualization*, which they define as the machine's ability to answer "questions about what a network—or parts of a network—are looking for by generating examples." In essence, feature visualization is the term we can use to describe both the process and the visual product of DeepDream's analysis of an image. As discussed above, DeepDream visualizes the network's interpretive associations of the various components of an image in its overall recognition process. It visualizes an entire layer of the network, yet other components of the network—a single neuron or a network channel, among other options—can also be visualized, giving more precise visual information about the recognition process. All of this can be called feature visualization. On the level of a neuron, this is somewhat comparable to observing how we come to recognize the visual development of one biological cell in a person, instead of the process of beholding that person in the entirety of his or her biological makeup.

Although the ramifications for our knowledge of the deep-learning process by using feature visualization is tremendous, its immediate application to art is not. Neither the vivid images produced by DeepDream nor the ability to replicate preexisting styles in painting onto new images have direct interest to art historians in a conventional sense, yet the *ability* to now do these things, and the implications of these sophisticated capabilities, will no doubt underscore this phase in the relationship of art and artificial intelligence as an historic turning point. It's unfortunate that the immediate applications of the very research that is actively building the development of advanced image analysis are showcased as a kind of visual entertainment, thus eclipsing the academic community most attuned to image analysis: art historians. Research projects that involve the collaboration of art historians, which include their assistance in the construction of the datasets used for training machines in image recognition and their involvement in how we interpret the image content produced in feature visualization, will help change this. While this may seem simple, what goes into the construction of a useful art historical dataset is not recognized as the major cultural heritage management and diplomatic feat that it should be, and getting art historians interested to study feature visualization is no easy task.

Perhaps this collaborative direction would lead to a machine's analysis of Stieglitz's photograph as "art" portraying Georgia O'Keeffe, and recognize the compositional strategy of the image as being indicative of the photographer's work, and date the photo (fig. 8).[39] Conceivably, it could even interpret the gesture of the artist holding her collar as giving symbolic meaning to the image in a manner that is sensitive to the cultural period in which it was produced, and bring attention to the other photographs that tie Stieglitz with O'Keeffe. Mostly based just on vision technology, the machine could interpret the clasping hands not as grasps at a mobile phone, but as the unique hands of an artist, informing the interpretation of her portrait. All of these observations could potentially be made without the machine's understanding of anything other than visual records, and already this would be a significant building block in our art historical understanding of the photograph. In actuality, Alfred Stieglitz took more than three hundred photographs of Georgia O'Keeffe between 1917 and 1937, and famously they shared a romantic relationship.[40] It was Stieglitz's intention to build a large composite portrait of O'Keeffe out of all the images. He believed that portraiture could reflect a person's life story and that photography captured but a moment from that narrative.[41]

Panofsky recognized that the immediate impetus to interpret an image upon its viewing is in fact so strong in the act of human visual perception that training is required to separate one's analysis of its formal characteristics over its subject. His memorable example is of an acquaintance greeting him on the street with the gesture of lifting his hat, and he describes the act of interpretation of this "event" with iconographical analysis that allows one to see how meaning is built into the visual interpretation of the scene. Ultimately, we learn of the "personality" of this acquaintance, a man of the early twentieth century engaging in a polite greeting, yet Panofsky draws attention to the role of inference in making this assessment. He writes in description of the intrinsic meaning of the image: "We could not construct a mental portrait of the man on the basis of this single action, but only by coordinating a large number of similar observations and by interpreting them in connection with our general information as to his period, nationality, class, intellectual traditions and so forth. Yet all the qualities which this mental portrait would show explicitly are implicitly inherent in every single action; so that, conversely, every single action can be interpreted in the light of those qualities."[42]

It is not coincidence that it is precisely this paradoxical theory of visual perception that is at the root of machine-based image interpretation using deep-learning techniques. However, if new research in neurobiology and cognitive science determines that the core components of visual perception operate in a less distillable manner than Panofsky suggests, this may be a false paradox. That his theories make notable accommodation for the instinct we have to bring interpretation to an image at the moment of its sighting draws attention to a more fundamental aspect of visual perception than is the focus of his methodology.[43] Current studies in *synesthesia*, the perceptual phenomenon in which the experience of one sense automatically and simultaneously triggers another, already provide compelling proof of the indivisibility of a visual experience with other interpretative processes, at least in some persons.[44] The ability of some people to have an auditory, or other sensory, sensation of what they are seeing in conjunction with their visual experience also demonstrates that visual perception is not a uniform human process, and this may have as much to do with the unique biological makeup of individuals as it does with the role of the environment in the shaping of our interpretative powers. Fortunately, research is under way in hearing-motion-related synesthesia, which may further validate the experiences of many noted artists and musicians that have brought attention to this phenomenon recently as well as in the past.[45]

    In the same vein, computer vision scientists may be independently discovering the unity of form and content through the exploration of their separateness, in a radically different way than even the modernists could have imagined, and questions about the image-recognition process remain. In many ways the idea of discovering the contents of the black box in the recognition process at large is just as elusive as the notion of the uniformity of visual experiences from person to person. If every image-recognition program that utilizes deep neural networks is unique in its formation of its interpretive capabilities, in relation to its training set, how can we continue to construct a general picture of this process without illustrating only different examples of it? One component of this theoretical quandary is addressed by "transfer learning," an area of research that actively seeks to replicate learned information in new programming conditions.[46]

Speaking at the Neural Information Processing Systems (NIPS) conference in 2016, Andrew Ng memorably predicted that after supervised learning, transfer learning would be the next driver in machine-learning commercial success.[47]

At the 2017 NIPS conference, the growing debate in computer science on the need for there to be more scientific inquiry regarding the mechanisms at work in the deep-learning process was underscored by computer scientist Ali Rahimi in his Test-of-Time Award presentation.[48] His criticism that more rigor is required for the study of this material before it is further expanded has already been sharply rebuked by the director of Facebook AI research and professor in the NYU School of Engineering Yann LeCun.[49] Already fostering much debate in the field of computer science, the divisiveness of this issue concerns the process of scientific discovery itself.[50] Should innovation in the age of AI be allowed to occur with a build-it-first engineering approach, or should we wait for theories and the rigorous analysis of processing models to guide development? This core question has much resonance with the study of human visual perception and art. Art historians don't have to utilize Panofsky's model of iconographical analysis to "do" art history, and many do not, but understanding how it is useful and where it may fail gives us much insight into the act of visual analysis itself.

Even though the most advanced AI in the field of vision technology today is mostly operating on what Panofsky termed the "pre-iconographical" level, we must recognize that a functional foundation for much higher-level vision tasks is quickly being built, whether it is fully understandable or not, and this is happening through models that are not replicative of, but analogous to, human visual processes, and by extension, to a basic level of art historical analysis.[51] Why not look to the fields within the umbrella category of neuroscience for methodological inspiration in art history, especially given the long history of the relationship between the arts and sciences? If we as a discipline revive research in iconography and formalism in collaboration with computer scientists for use on this new frontier of art and AI, we could help shape the future navigation and interpretation of all digital images, and this would have a profound impact on the course of the history of art history, and visual culture at large.[52] In sum, the development and use of deep neural networks afford us the opportunity to write the next chapter in our "studies in iconology," iconology in the age of AI.

________________

This paper was originally presented as *Dream Formulations and Image Recognition: Algorithms for the Study of Renaissance Art*, at *Critical Approaches to Digital Art History*, The Villa I Tatti, The Harvard University Center for Italian Renaissance Studies and The Newberry Center for Renaissance Studies, Renaissance Society of America Annual Meeting, Chicago, 31 March 2017. This paper would not have been possible without the research developments of Google Scientist Alexander Mordvintsev and his colleagues. I am also very grateful for Alexander's collaboration on this article with me. Additionally, Scott Dexter, Professor of Computer and Information Science at CUNY Graduate Center, provided me with feedback on the public presentation-version of this research, and Denis Cummings and Sharon Herson gave me invaluable help in the editing process. Lastly, I thank Angela Dressen for her organization of the RSA session and her invitation to publish this research.

[1] Alexander Mordvintsev, Christopher Olah, and Mike Tyka, *Inceptionism: Going Deeper into Neural Networks*, in: Google Research Blog, https://research.googleblog.com/2015/06/inceptionism-going-deeper-into-neural.html, 17-06-2015.

[2] The media focused on the use of the algorithm as an artistic tool. See Matt McFarland, *Google's psychedelic 'paint brush' raises the oldest question in art*, in: The Washington Post, https://www.washingtonpost.com/news/innovations/wp/2016/03/10/googles-psychedelic-paint-brush-raises-the-oldest-question-in-art, 10-03-2016.

[3] The term "neural networks" is a more popular example of this trend. By contrast, it is culturally significant to note that computerizing metaphors are also now being used to describe the human condition. Max Tegmark has recently categorized the evolution of the human species according to the relationship of our "hardware" and "software" in 1.0-, 2.0-, and 3.0-based versions. See Max Tegmark, *Life 3.0: Being Human in the Age of Artificial Intelligence*, New York 2017.

[4] The program was first named "inceptionism" in reference to Christopher Nolan's 2010 film *Inception*.

[5] The perceptual capabilities of machines have expanded greatly on account of deep neural networks. See Alex Krizhevsky, Ilya Sutskever, and Geoffrey E. Hinton, *Imagenet classification with deep convolutional neural networks*, in: Proceedings of the 25th International Conference on Neural Information Processing Systems, vol. 1, 2012, p. 1097-1105; Yang Taigman et al., *Deepface: Closing the gap to human-level performance in face verification*, in: 2014 IEEE Conference on Computer Vision and Pattern Recognition, 2014, p. 1701-1708.

[6] Some examples of the uses of deep neural networks to produce artistic imagery are discussed in Leon A. Gatys, Alexander S. Ecker, and Matthias Bethge, *A Neural Algorithm of Artistic Style*, in: arXiv:1508.06576v2 (cs.CV), https://arxiv.org/abs/1508.06576, 02-09-2015; Ahmed Elgammal et al., *CAN: Creative Adversarial Networks Generating 'Art' by Learning About Styles and Deviating from Style Norms*, arXiv:1706.07068 [cs.AI], https://arxiv.org/abs/1706.07068, 21-06-2017. Also, "Unhuman: Art in the Age of AI" was the first exhibition of wholly algorithmically produced art using deep-learning techniques by the AICAN algorithm. It was curated by the author and was exhibited in Los Angeles, CA, and Frankfurt, Germany, in October 2017.

[7] Self-driving cars and automated surgeries exemplify this trend.

[8] The observation that developments in computer vision technology have often been sequestered to the fields of specialists, yet the application of research in this area has increasingly been used by artists was already made by Golan Levin in 2006. See Golan Levin, *Computer Vision for Artists and Designers: Pedagogic Tools and Techniques for Novice Programmers*, in: Journal of Artificial Intelligence and Society, vol. 20, issue 4, September 2006, p. 462–482.

[9] This statement is a direct reference to art historian Erwin Panofsky's pioneering work on the study of *iconography*, which he terms *iconology*. The title of this article also makes reference to Panofsky's contributions to art history in this area. The "machine-learned" image is a reference to the field of machine learning in computer science that is sometimes used synonymously with the term "AI," although the concept of artificial intelligence lacks definitional precision and has a long and problematic history unrelated to machine learning.

[10] Olga Russakovsky et al., *ImageNet Large Scale Visual Recognition Challenge,* in: International Journal of Computer Vision, vol. 115, issue 3, December 2015, p. 211-252.

[11] Aaron M. Borenstein, *Is Artificial Intelligence Permanently Inscrutable?* in: Nautilus, issue 40, http://nautil.us/issue/40/learning/is-artificial-intelligence-permanently-inscrutable, 01-09-2016.

[12] Ibid.

[13] Mordvintsev, Olah, and Tyka 2015, *Inceptionism: Going Deeper into Neural Networks*.

[14] Ibid.

[15] Ramprasaath R. Selvaraju et al., *Grad-CAM: Visual Explanations from Deep Networks via Gradient-based Localization*, in: arXiv:1610.02391v3 [cs.CV], https://pdfs.semanticscholar.org/5582/bebed97947a41e3ddd9bd1f284b73f1648c2.pdf, 21-03-2017.

[16] This is occurring under the auspices of the museum's Photoarchive department.

[17] The program is user friendly and can easily be accessed for experimentation at http://gradcam.cloudcv.org/classification and http://gradcam.cloudcv.org/captioning.

[18] The analyzed image in discussion is the *Madonna and Child*, attributed to Berlinghiero (active by 1228, died by 1236), tempera on wood, gold ground, at The Metropolitan Museum of Art.

[19] The analyzed image in discussion is *Manuscript leaves fragment* from Byzantine Egypt, 4th–7th century, ink on parchment, at The Metropolitan Museum of Art.

[20] The image was in fact classified differently than its caption would have suggested, recognizing the top of the collar as the significant part of the image and thus predicting the label "neck brace."

[21] It is likely that our understanding of this phenomenon will change in the near future.

[22] As specific approaches in AI are improved and may even reach a level that scientists have termed AGI (Artificial General Intelligence), this will no longer be an issue. See Hans Moravec, *When Will Computer Hardware Match the Human Brain?* in: Journal of Evolution and Technology, vol. 1, 1998.

[23] See Rudolf Arnheim, *Art and Visual Perception: A Psychology of the Creative Eye*, Berkeley 1954; Michael Baxandall, *Patterns of Intention: On the Historical Explanation of Pictures*, New Haven 1985, (especially chapter 5, p. 70-104).

[24] Eleanor H. Rosch, *Natural Categories*, in: Cognitive Psychology, vol. 4, issue 3, 1973, p. 328-350; Jean Piaget, *The Construction of Reality in the Child*, New York 1955.

[25] Eleanor H. Rosch, *Cognitive Reference Points*, in: Cognitive Psychology, vol. 7, issue 4, 1975, p. 532-547.

[26] In future research it would be interesting to philosophically compare prototype theory to Plato's theory of forms. I thank Sharon Herson for pointing out this connection.

[27] Eleanor H. Rosch, *Cognitive Representations of Semantic Categories,* in: Journal of Experimental Psychology: General, vol. 104, issue 3, September 1975, p. 192-233.

[28] Jean Piaget, *The Child's Conception of the World*, London 1929. Also see Susan A. Gelman, *The Essential Child: Origins of Essentialism in Everyday Thought*, Oxford 2003.

[29] Mordvintsev, Olah, and Tyka 2015, *Inceptionism: Going Deeper into Neural Networks*.

[30] Erwin Panofsky, *Studies in Iconology: Humanistic Themes in the Art of the Renaissance,* New York 1939.

[31] Ibid.

[32] Michael Ann Holly, *Panofsky and the Foundations of Art History*, Ithaca, NY, 1985, p. 114-157.

[33] Gatys, Ecker, and Bethge 2015, *A Neural Algorithm of Artistic Style*, p. 9.

[34] Alexander Mordvintsev, Christopher Olah, and Mike Tyka, *DeepDream - a code example for visualizing Neural Networks*, in: Google Research Blog, https://research.googleblog.com/2015/07/deepdream-code-example-for-visualizing.html, 01-07-2015.

[35] On 8 March 2016, Washington Post Innovations editor Matt McFarland made this comment while interviewing the author on the use of the DeepDream algorithm to create art.

[36] Grant Taylor, *When the Machine Made Art, The Troubled History of Computer Art*, New York 2014.

[37] Chris Olah, Alexander Mordvintsev, and Ludwig Schubert, *Feature Visualization: How neural networks build up their understanding of images*, in: Distill, https://distill.pub/2017/feature-visualization, 07-11-2017.

[38] Since the presentation of my research in this area at RSA 2017, the advances in computer vision technology have expanded so remarkably that critical additions needed to be added to this publication, as the presentation version of it had already became partially outdated in the publication process. This phenomenon underscores one inherent difficulty in the pursuit of collaborative projects that cross the arts and sciences.

[39] Ahmed Elgammal et al. at the Rutgers Art and Artificial Intelligence Laboratory have been able to discriminate images of "art" from other images using deep-learning techniques. See Ahmed Elgammal et al. 2017, *CAN: Creative Adversarial Networks Generating 'Art' by Learning About Styles and Deviating from Style Norms*; Kanako Abe, Babak Saleh, and Ahmed Elgammal, *An Early Framework for Determining Artistic Influence*, in: New Trends in Image Analysis and Processing – ICIAP 2013, September 2013, p. 198-207.

[40] For an overview of Georgia O'Keeffe's life and her relation with Alfred Steiglitz, see Lisa Messinger. *Georgia O'Keeffe (1887–1986)*, in: Heilbrunn Timeline of Art History. New York: The Metropolitan Museum of Art, http://www.metmuseum.org/toah/hd/geok/hd_geok.htm, 2004.

[41] See Daniel Malcolm, *Stieglitz, Steichen, Strand: Masterworks from the Metropolitan Museum of Art*, New Haven 2010, p. 22.

[42] Erwin Panofsky, *Iconography and Iconology: An Introduction to The Study of Renaissance Art*, in: Meaning in the Visual Arts, Garden City, NY, 1972, p. 26-64.

[43] In his example of the man gesturing with his hat, Panofsky calls the automatic interpretation that occurs upon viewing as having "overstepped the limits of purely formal perception and (having) entered a first sphere of subject matter or meaning." He later critiques "formal analysis," as developed by Heinrich Wölfflin, as even needing to avoid "expressions as 'man,' 'horse,' or 'column,'" which already demonstrate interpretability on the level of the object. Ibid., p. 26 and p. 30.

[44] See Richard E. Cytowic, *Synesthesia: A Union of the Senses,* Cambridge 2002.

[45] The doctoral thesis by Christopher Fassnidge, "Exploring the prevalence and underlying mechanisms of Hearing-Motion Synesthesia with psychophysics, EEG, and neurostimulation," is underway in the Department of Psychology and the Cognitive Neuroscience Research Unit at the City University of London. David Hockney, Wassily Kandinsky, and Duke Ellington are but a few famous *synesthetes*.

[46] For an overview of transfer learning, see Sebastian Ruder, *Transfer Learning—Machine Learning's Next Frontier*, in: ruder.io, http://ruder.io/transfer-learning, 21-03-2017.

[47] Andrew Y. Ng, *Nuts and Bolts of Building Applications using Deep Learning*, Thirtieth Annual Conference on Neural Information Processing Systems (NIPS), Barcelona, 5-10 December 2016.

[48] Ali Rahimi, *Test of Time Award Presentation*, Thirty-first Annual Conference on Neural Information Processing Systems (NIPS), Long Beach, CA, 4-9 December 2017.

[49] Yann LeCun, *My take on Ali Rahimi's "Test of Time" award talk at NIPS.*, in: Facebook, https://www.facebook.com/yann.lecun/posts/10154938130592143, 06-12-2017.

[50] I thank Eliezer Bernart, Ph.D. candidate in Computer Science at Federal University of Rio Grande do Sul, for drawing my attention to the debate that has already been generated by Ali Rahimi's presentation and Yann LeCun's rebuttal. Ferenc Huszar's blog post is an example of this discussion. See Ferenc Huszar, *Alchemy, Rigour, and Engineering*, in: inFERENCe, http://www.inference.vc/my-thoughts-on-alchemy, 07-12-2017. More debate on this subject is expected.

[51] Emily L. Spratt and Ahmed Elgammal, *Computational Beauty: Aesthetic Judgment at the Intersection of Art and Science*, in: Computer Vision - ECCV 2014 Workshops: Zurich, Switzerland, September 6-7 and 12, 2014, Proceedings, Part I, 2015.

[52] To this end, I am organizing with my colleagues Louisa Wood Ruby and Ellen Prokop the symposium *Searching Through Seeing: Optimizing Vision Technology for the Arts*, which will take place at the Frick Collection in New York on 12-13 April 2018. The aim of the symposium is to better develop image recognition and retrieval technologies for use in art history.

**Figures**

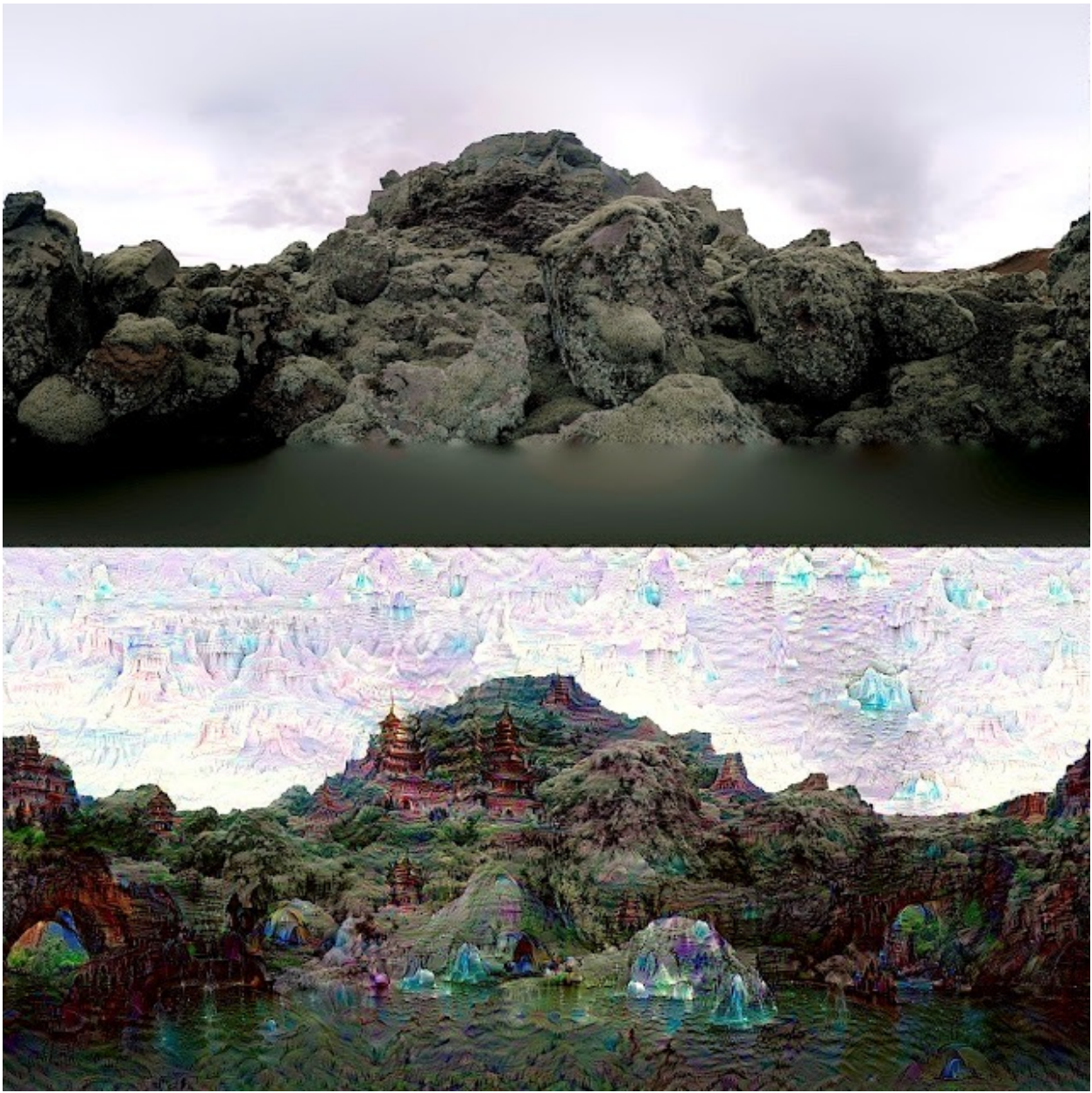

Fig. 1 DeepDream; Visualization Input (top) and Output (bottom); 2015; Digital Image. (Google Research Blog, with permission from Google)

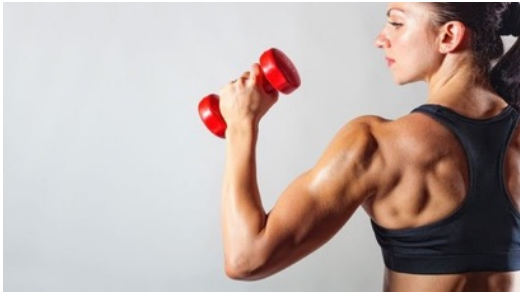

Fig. 2 Stock image of weightlifter curling a dumbbell; Digital Image.

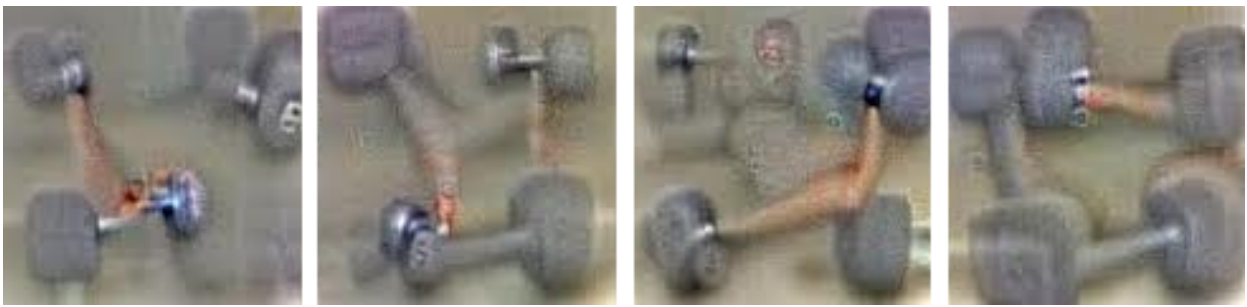

Fig. 3 DeepDream; Visualization of Dumbbells with Attached Arms; 2015; Digital Image. (Google Research Blog, with permission from Google)

## Result of Grad-CAM for Classification

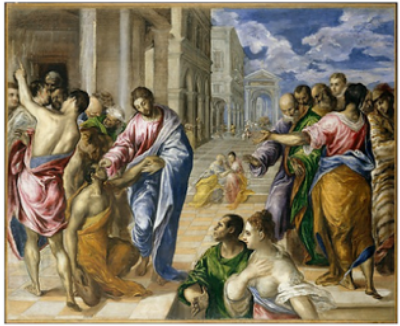

Predicted Label : carousel

Generating Grad-CAM visualizations for: carousel

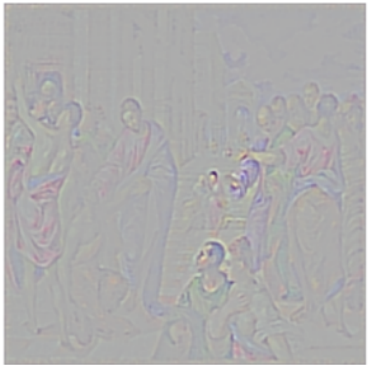
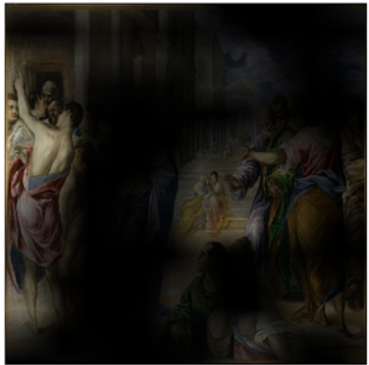
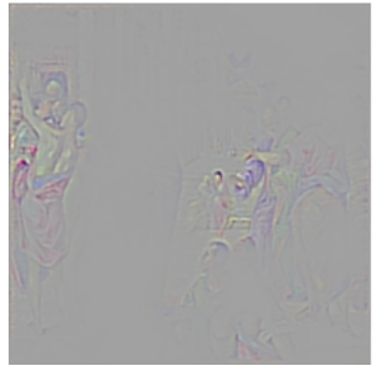

Guided Backprop — Grad-CAM — Guided Grad-CAM

## Credits

Built by @deshraj

Fig. 4 Grad-CAM demo results for El Greco, *Christ Healing the Blind*, The Metropolitan Museum of Art, New York.

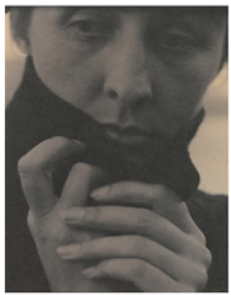
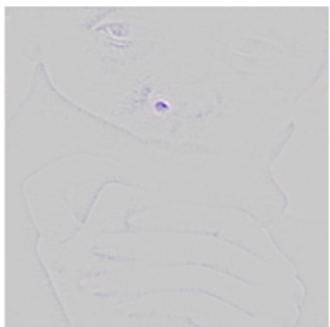
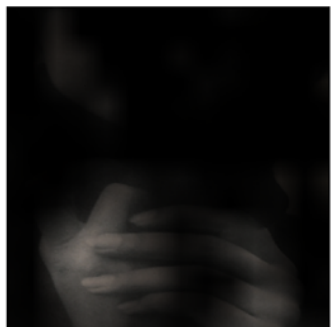
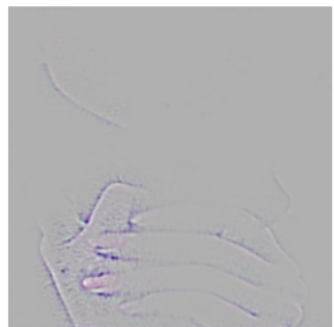

Fig. 5 Grad-CAM demo results for Alfred Stieglitz, *Georgia O'Keeffe*, The Metropolitan Museum of Art, New York.

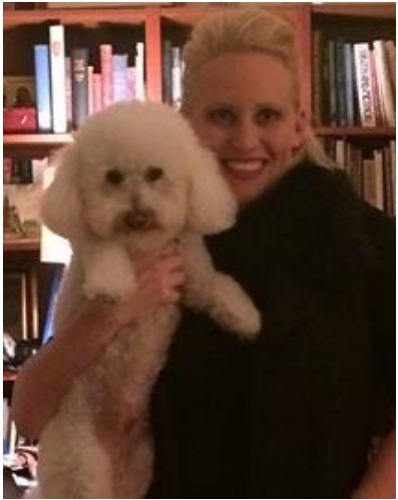

Fig. 6 Pumba, Bichon Frise; 2017. (Author's Photograph)

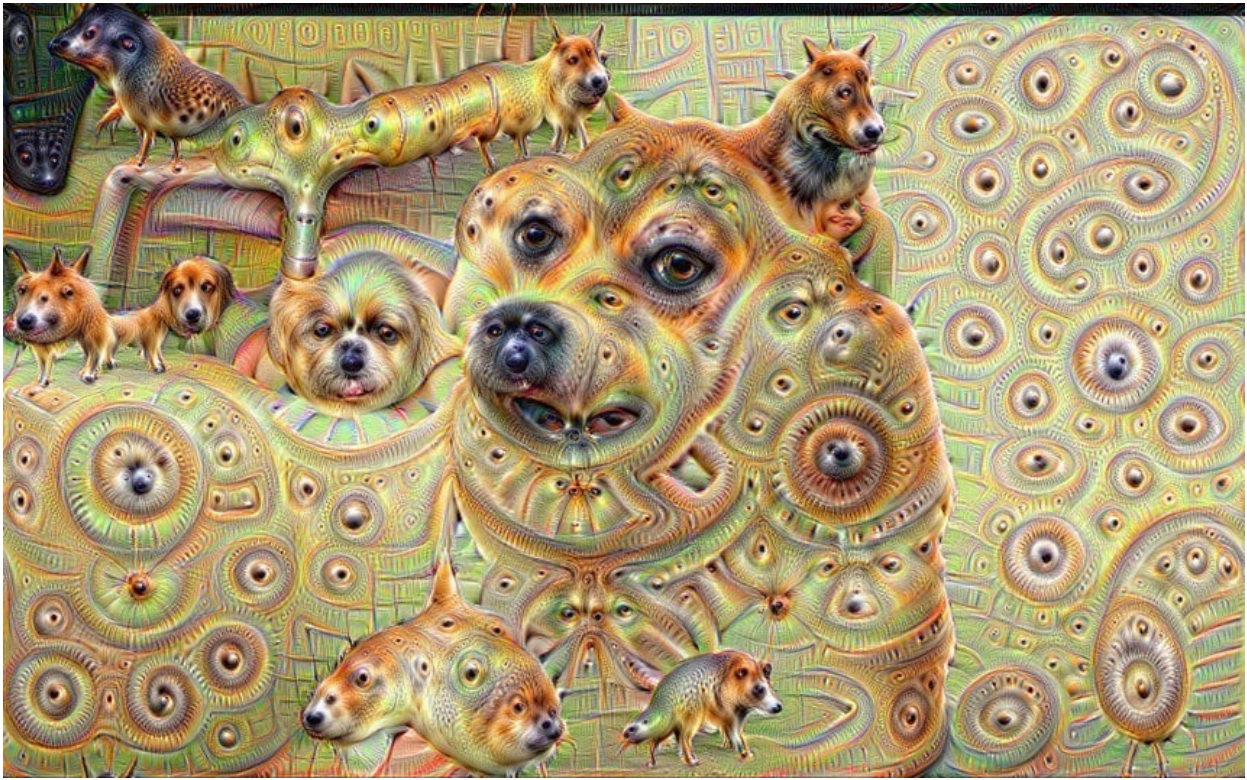

Fig. 7 DeepDream; Dog Visualization; 2015; Digital Image. (With permission from Google)

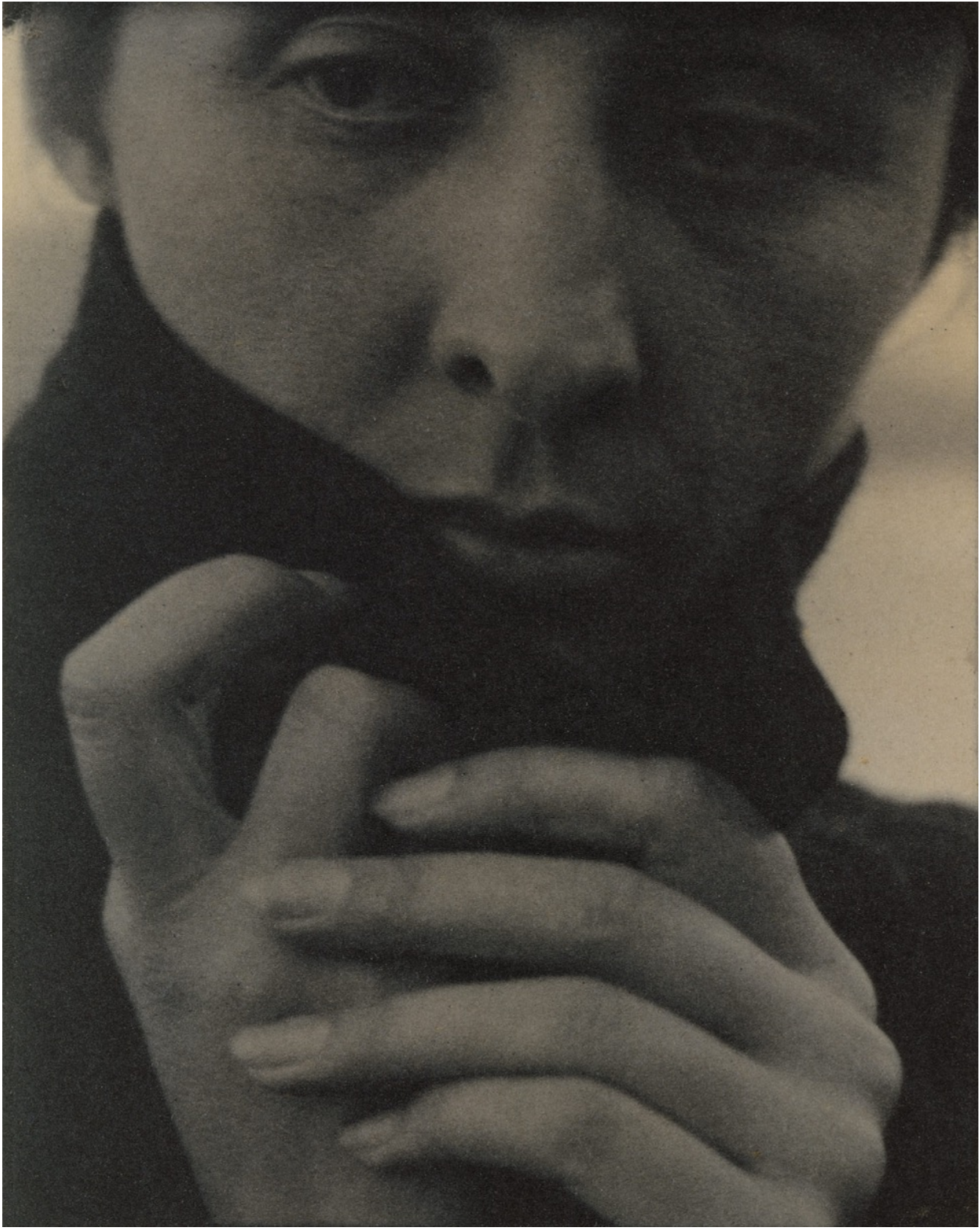

Fig. 8 Alfred Stieglitz; *Georgia O'Keeffe*; 1918; Palladium Print, Photograph; 11.7 x 9 cm; The Metropolitan Museum of Art, New York.